\documentclass{PoS}  

\usepackage{amsmath}

\def\BA{\begin{eqnarray}}
\def\EA{\end{eqnarray}}
\def\BAN{\begin{eqnarray*}}
\def\EAN{\end{eqnarray*}}

\def\beq{\begin{equation}}
\def\eeq{\end{equation}}
\def\bea{\begin{eqnarray}}
\def\eea{\end{eqnarray}}
\def\g5{\gamma_5}

\def\tr{\mathrm{tr}}

\PoS{PoS(LAT2005)041}  
\title{Pseudoscalar Mass and Decay Constant in Lattice QCD 
with Exact Chiral Symmetry\thanks{  
This work was supported in part by the National Science Council (ROC), 
under Grant No. NSC93-2112-M002-016, and the
National Center for High Performance Computation at Hsinchu.}}
\ShortTitle{Pseudoscalar Mass and Decay Constant in Lattice QCD}  
\author{\speaker{Ting-Wai Chiu}, Tung-Han Hsieh, Jon-Yu Lee, Pei-Hua Liu, Hsiu-Ju Chang \\ 
Physics Department, National Taiwan University, Taipei, Taiwan 106, Taiwan \\ 
E-mail: \email{twchiu@phys.ntu.edu.tw} }  

\abstract{
The masses and decay constants of pseudoscalar mesons
$ D $, $ D_s $, and $ K $ are determined 
in quenched lattice QCD with exact chiral symmetry.
For 100 gauge configurations generated with single-plaquette
action at $ \beta = 6.1 $ on the $ 20^3 \times 40 $ lattice,
we compute point-to-point quark propagators for 30 quark masses
in the range $ 0.03 \le m_q a \le 0.80 $, and measure the
time-correlation functions of pseudoscalar and vector mesons.
The inverse lattice spacing $ a^{-1} $
is determined with the experimental input of $ f_\pi $,
while the strange quark bare mass ($ m_s a = 0.08 $), and the charm
quark bare mass ($ m_c a = 0.80 $) are fixed such that the masses of
the corresponding vector mesons are in good agreement with
$ \phi(1020) $ and $ J/\psi(3097) $ respectively.
Our results of pseudoscalar-meson decay constant are: 
$ f_K = 152(6)(10) $ MeV, $ f_D = 235(8)(14)$ MeV,
and $ f_{D_s} = 266(10)(18) $ MeV \cite{Chiu:2005ue}. 
The latest experimental result of $ f_{D^+} $ from CLEO \cite{Artuso:2005ym} 
is in good agreement with our prediction.
} 

\FullConference{XXIIIrd International Symposium on Lattice Field Theory\\ 
25-30 July 2005\\ 		 
Trinity College, Dublin, Ireland}  

\begin{document}  

\section{Introduction}

The pseudoscalar-meson decay constants
(e.g., $ f_D $, $ f_{D_s} $, $ f_B $ and $ f_{B_s} $)
play an important role in extracting the CKM matrix elements
(e.g., the leptonic decay width of $ D_s^+ \to l^+ \nu_l $ is
proportional to $ f_{D_s}^2 | V_{cs} |^2 $),
which are crucial for testing the flavor sector of the
standard model via the unitarity of CKM matrix.
Experimentally, precise determination of $ f_{D} $
and $ f_{D_s} $ will result from the high-statistics program
of CLEO-c, however, the determination of $ f_{B} $ and $ f_{B_s} $
remains beyond the reach of current experiments.

Theoretically, lattice QCD provides a solid framework to compute
the masses and decay constants of pseudoscalar mesons
(as well as other physical observables) nonperturbatively
from the first principles of QCD.
Thus reliable lattice QCD determinations of $ f_B $ and $ f_{B_s} $
are of fundamental importance, in view of their experimental
determinations are still lacking. Obviously, the first step
for lattice QCD is to check whether the lattice determinations of
$ f_D $ and $ f_{D_s} $ will agree with those coming from
the high-statistics charm program of CLEO-c. This motivated
our study in Ref. \cite{Chiu:2005ue}. It turns out that our 
{\it predictive} value of $ f_D = 235(8)(14) $ MeV (posted on June 26) 
is in good agreement with the experimental result 
$f_{D^+}=(223 \pm 16^{+7}_{-9})~{\rm MeV} $
announced by CLEO-c at Lepton-Photon Symposium (LP2005) on July 1
\cite{Artuso:2005LP}. (Note that the latest value of $ f_{D^+} $ 
from CLEO-c has been updated to 
$f_{D^+}=(222.6\pm 16.7^{+2.8}_{-3.4})~{\rm MeV} $  
\cite{Artuso:2005ym}.)

In this talk, we review our results reported   
in Ref. \cite{Chiu:2005ue}, and also present our results for 
the ratios $ f_K/f_\pi $ and $ f_{D_s}/f_D $, which may be 
of interest from several viewpoints.      

Here we briefly outline our computations, and refer to 
Ref. \cite{Chiu:2005ue} (and references therein) for further details.  
First, we compute quenched quark propagators for 30 quark masses 
in the range $ 0.03 \le m_q a \le 0.80 $, in the framework of 
optimal domain-wall fermion proposed by Chiu \cite{Chiu:2002ir}.
Then we determine the inverse lattice spacing $ a^{-1} = 2.237(75) $ GeV
from the pion time-correlation function, with the experimental input
of pion decay constant $ f_\pi = 131 $ MeV.
The strange quark bare mass $ m_s a = 0.08 $
and the charm quark bare mass $ m_c a = 0.80 $ are fixed
such that the corresponding masses
extracted from the vector meson correlation
function agree with
$ \phi(1020) $ and $ J/\psi (3097) $ respectively.
Then the masses and decay constants of any hadrons
containing $ c, s $, and $ u (d) $ quarks\footnote{In this paper, we work in
the isospin limit $ m_u = m_d $.}
are predictions of QCD from the first principles,
with the understanding that
chiral extrapolation to physical $ m_{u,d} \simeq m_s/25 $
(or equivalently $ m_\pi = 135 $ MeV) is required
for any observables containing $ u(d) $ quarks.

The observable we measure is the time-correlation function for 
pseudoscalar meson ($ \bar q Q $)
\bea
\label{eq:CPS}
C_{P} (t) = \left<
\sum_{\vec{x}}
\tr\{ \gamma_5 (D_c + m_Q)^{-1}_{x,0} \gamma_5 (D_c + m_q)^{-1}_{0,x} \}
\right>_U   
\eea
where the subscript $ U $ denotes averaging over gauge configurations.
Here $ C_{P}(t) $ is measured for the following three categories: 
(i) Symmetric masses $ m_Q = m_q $,
(ii) Asymmetric masses with fixed $ m_Q = m_s = 0.08 a^{-1} $,
(iii) Asymmetric masses with fixed $ m_Q = m_c = 0.80 a^{-1} $, 
where $ m_q $ is varied for 30 masses in the range
$ 0.03 \le m_q a \le 0.80 $.
Note that for mesons composed of strange and/or charm quarks,
their masses and decay constants can be measured directly
without chiral extrapolation.

The decay constant $ f_P $ for a charged pseudoscalar meson $ P $ is 
defined by
\BAN
\left<0| A_\mu(0) | P(\vec{q}) \right> = f_P q_\mu
\EAN
where $ A_\mu = \bar q \gamma_\mu \gamma_5 Q $ is the axial-vector
part of the charged weak current after a CKM matrix element $ V_{qq'} $
has been removed.
Using the formula $ \partial_\mu A_\mu = (m_q + m_Q) \bar q \gamma_5 Q $,
one obtains
\bea
\label{eq:fP}
f_P = (m_q + m_Q )
\frac{| \langle 0| \bar q \gamma_5 Q | P(\vec{0}) \rangle |}{m_P^2}
\eea
where the pseudoscalar mass $ m_P a $ and the decay amplitude
$ z \equiv | \langle 0| \bar q \gamma_5 Q | P(\vec{0}) \rangle | $ 
can be obtained by fitting the pseudoscalar time-correlation function 
$ C_P(t) $ to the usual formula 
\bea
\label{eq:Gt_fit}
\frac{z^2}{2 m_P a} [ e^{-m_P a t} + e^{- m_P a(T-t)} ]
\eea

The outline of this paper is as follows. 
We present our
results of $ m_K $ and $ f_K $ in section 2, 
$m_D $, $ m_{D_s} $, $ f_D $, and $ f_{D_s} $ in section 3, 
and the ratios $ f_K/f_\pi $ and $ f_{D_s}/f_D $ in section 4. 
In section 5, we summarize and conclude with some remarks.

\section{$f_K$ and $ m_K $}

\begin{figure}[th]
\begin{center}
\begin{tabular}{@{}c@{}c@{ }c@{}c@{}}
\parbox[b][7.5cm][t]{6.5mm}{(a)} &
\parbox[b][7.5cm][t]{6.5cm}{
        \includegraphics*[height=7cm,width=6cm]{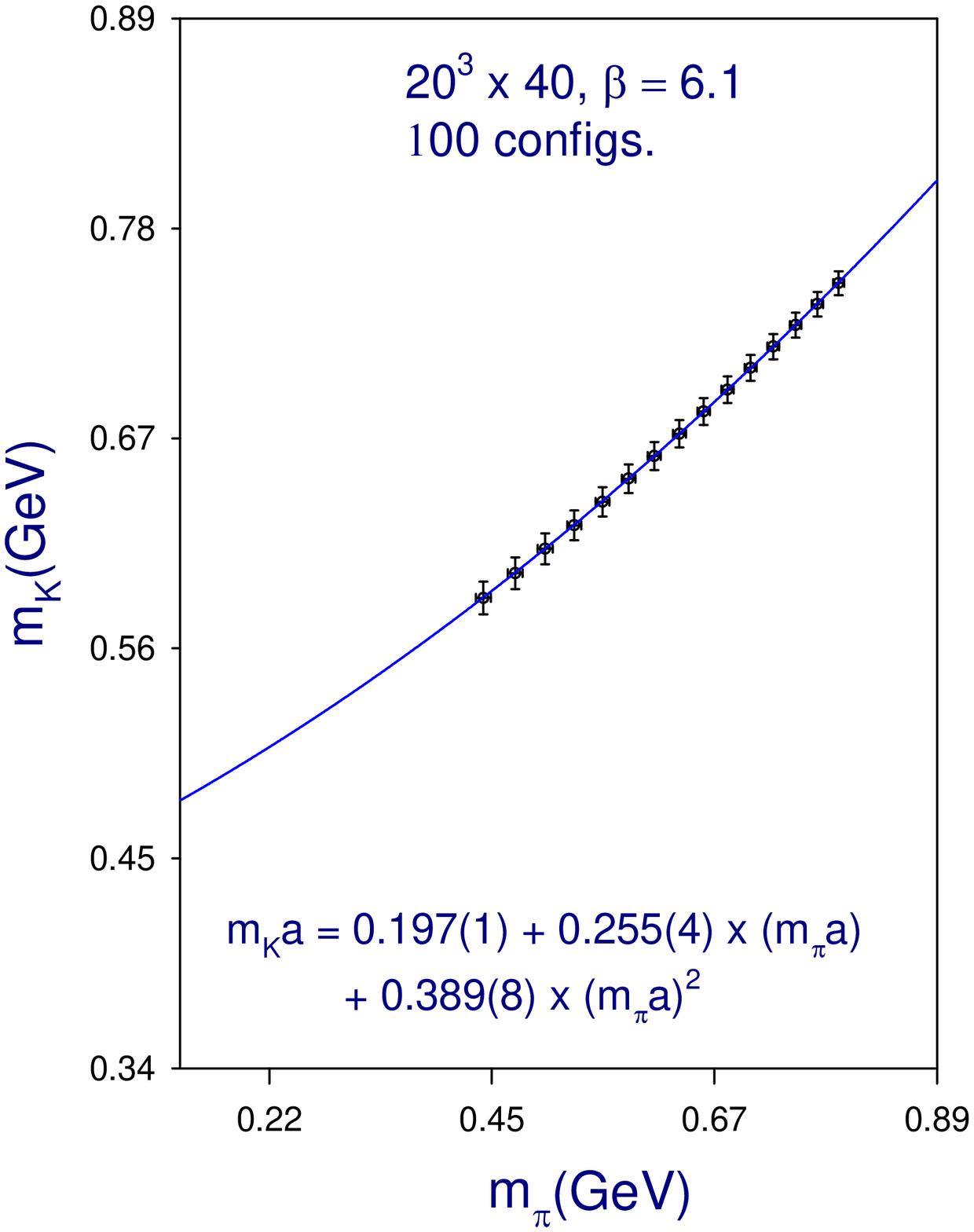}} &
\parbox[b][7.5cm][t]{6.5mm}{(b)} &
\parbox[b][7.5cm][t]{6.5cm}{
        \includegraphics*[height=7cm,width=6cm]{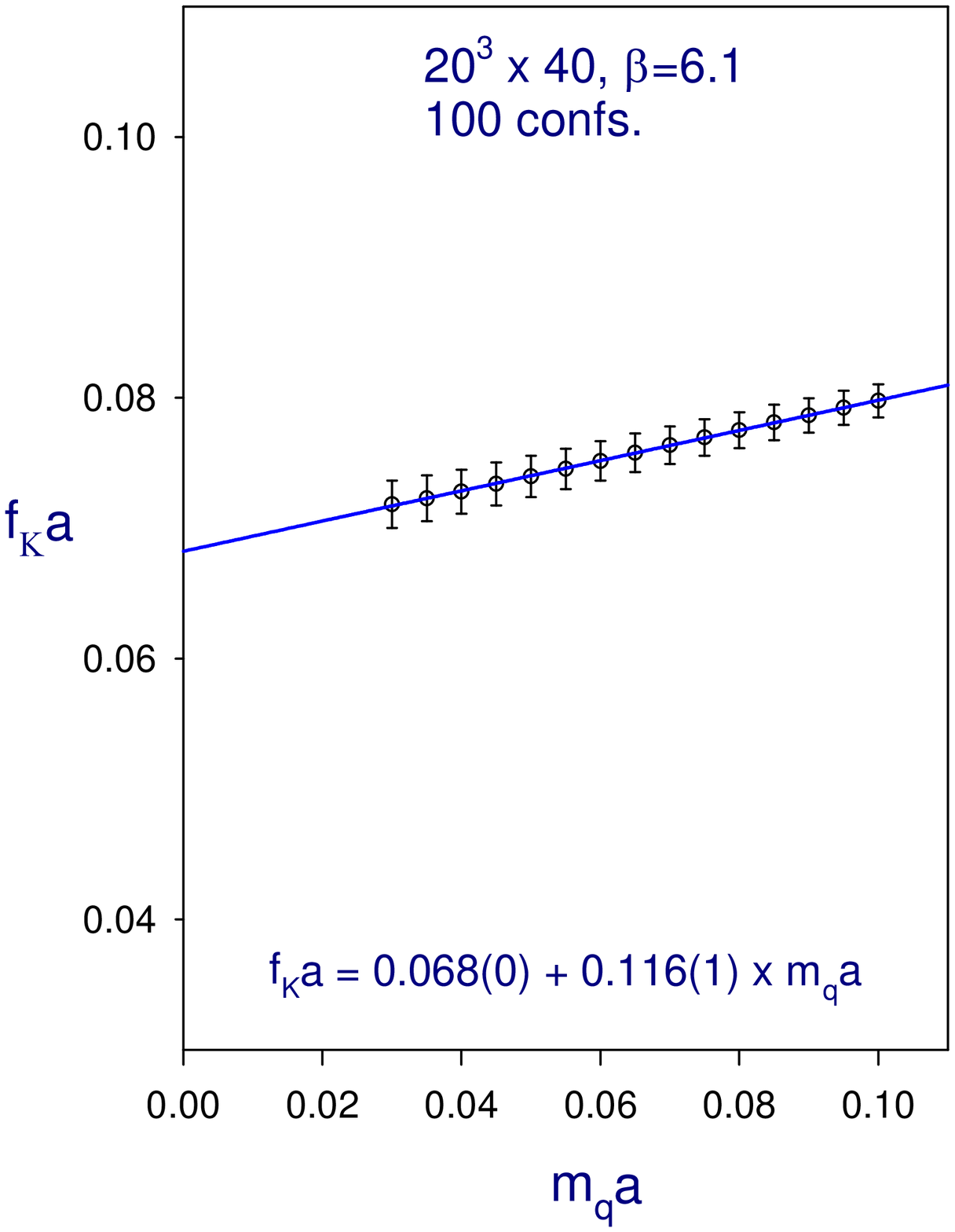}}
\end{tabular}
\vskip -0.8truecm
\caption{(a) 
The kaon mass $ m_K $ versus the pion mass $ m_{\pi} $ 
for 15 bare quark masses in the range $ 0.03 \le m_q a \le 0.10 $.
The solid line is the quadratic fit.
(b) The kaon decay constant $ f_K a $  
versus the bare quark mass $ m_q a $. The solid line is the linear fit.}
\label{fig:mK_fK}
\end{center}
\end{figure}

We measure the time-correlation function of kaon
$ C_{K}(t) $ (\ref{eq:CPS}) with $ m_Q $ fixed at 
$ m_s = 0.08 a^{-1} $, while $ m_q $
is varied for 30 masses in the range $ 0.03 \le m_q a \le 0.80$.
Then the data of $ C_K(t) $ 
is fitted by the formula (\ref{eq:Gt_fit})
to extract the kaon mass $ m_{K} a $ and the kaon decay constant
$ f_K a $.

In Fig. \ref{fig:mK_fK}a, the kaon mass $ m_K $ is plotted versus $ m_\pi $, 
for 15 quark masses in the range $ 0.03 \le m_q a \le 0.10 $.  
The data of $ m_K a $ can be fitted by  
$ m_K a = 0.197(1) + 0.255(4) (m_\pi a) + 0.389(8) (m_\pi a)^2 $.
At the physical limit $ m_\pi = 135 $ MeV, it gives 
$ m_K = 478(16) $ MeV, in good agreement with the experimental value 
of kaon mass ($ 495 $ MeV).

In Fig. \ref{fig:mK_fK}b, $ f_K a $ is plotted versus 
bare quark mass $ m_q a $. 
The data is well fitted by the straight line  
$ f_{K} a = 0.068(0) + 0.116(1)  \times  (m_q a) $.
At $ m_q a = 0 $, it gives $ f_K = 152(6) $ MeV, in agreement
with the value $ f_{K^+} = 159.8 \pm 1.4 \pm 0.44 $ MeV
complied by PDG \cite{Eidelman:2004wy}.

\section{$f_D$, $ f_{D_s} $, $m_D$, and $ m_{D_s} $}

\begin{figure}[th]
\begin{center}
\begin{tabular}{@{}c@{}c@{ }c@{}c@{}}
\parbox[b][7.5cm][t]{6.5mm}{(a)} &
\parbox[b][7.5cm][t]{6.5cm}{
        \includegraphics*[height=7cm,width=6cm]{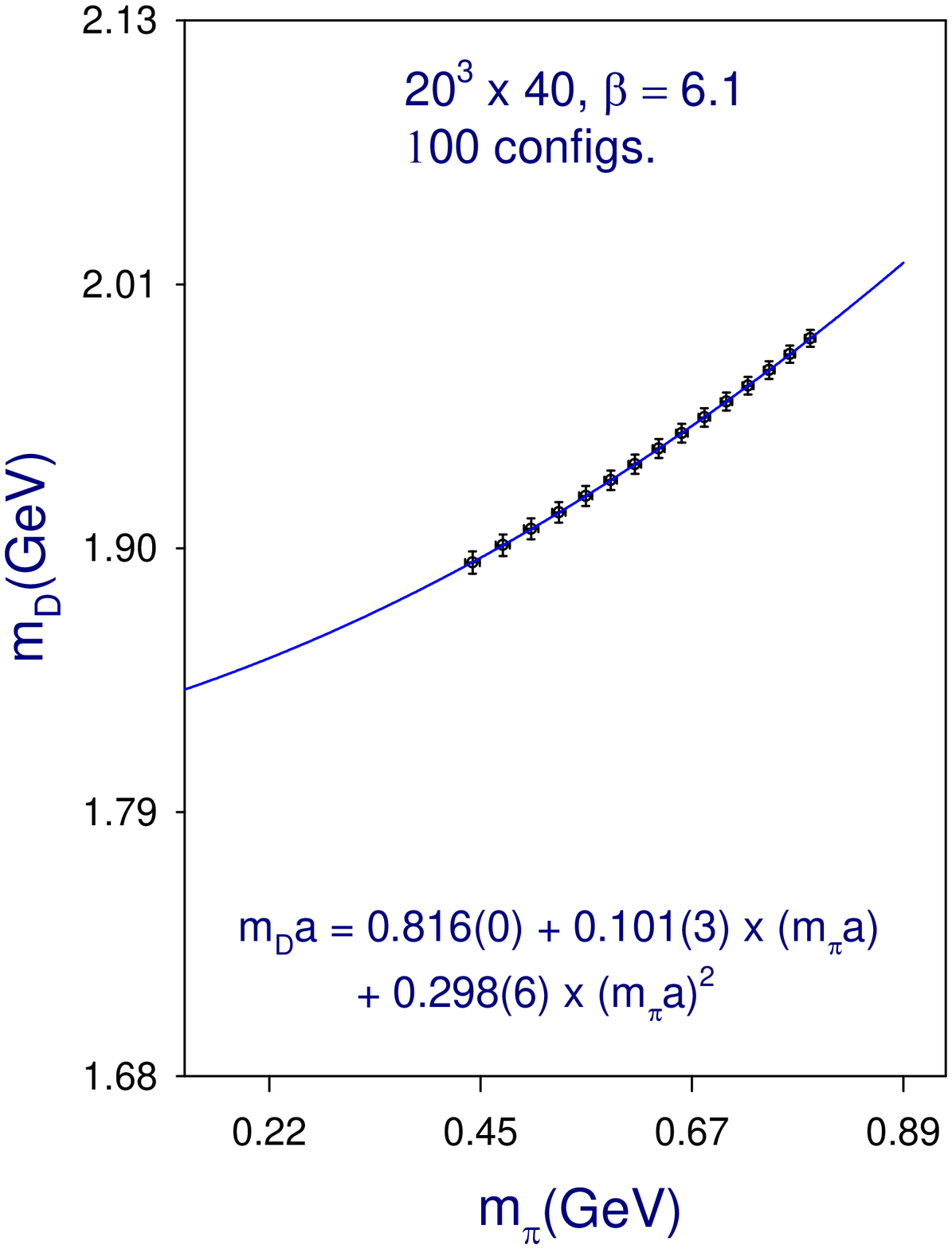}} &
\parbox[b][7.5cm][t]{6.5mm}{(b)} &
\parbox[b][7.5cm][t]{6.5cm}{
        \includegraphics*[height=7cm,width=6cm]{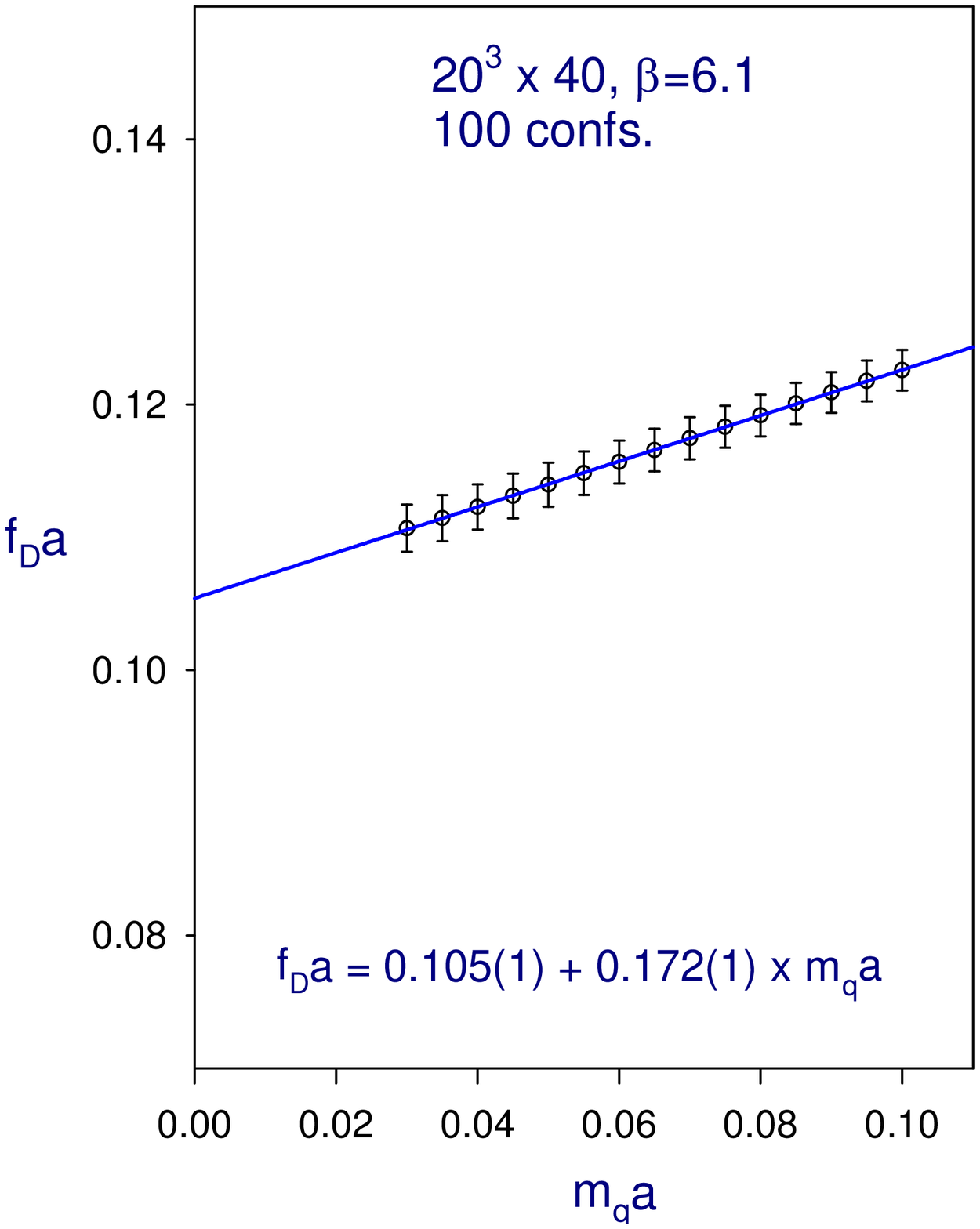}}
\end{tabular}
\vskip -0.8truecm
\caption{(a) 
The $ D $-meson mass $ m_D $ versus the pion mass $ m_{\pi} $ 
for 15 bare quark masses in the range $ 0.03 \le m_q a \le 0.10 $.
The solid line is the quadratic fit.
(b) The D-meson decay constant $ f_D a $  
versus the bare quark mass $ m_q a $. The solid line is the linear fit.}
\label{fig:mD_fD}
\end{center}
\end{figure}

Now we turn to charmed pseudoscalar mesons.  
We measure the time-correlation function  
$ C_{D}(t) $ (\ref{eq:CPS}) with $ m_Q $ fixed at 
$ m_c = 0.80 a^{-1} $, while $ m_q $
is varied for 30 different masses in the range $ 0.03 \le m_q a \le 0.80$.
Then the data of $ C_D(t) $ 
is fitted by the formula (\ref{eq:Gt_fit})
to extract the mass $ m_{D} a $ and decay constant $ f_D a $. 

In Fig. \ref{fig:mD_fD}a,    
$ m_D a $  is plotted versus $ m_\pi a $, 
for 15 quark masses in the range $ 0.03 \le m_q a \le 0.10 $. 
The data of $ m_D a $ can be fitted by 
$ m_D a = 0.816(0) + 0.101(3) (m_\pi a) + 0.298(6) (m_\pi a)^2 $. 
At $ m_\pi = 135 $ MeV, it gives $ m_D = 1842(15) $ MeV, 
in good agreement with the mass of $ D $ meson ($ 1865 $ MeV).
In Fig. \ref{fig:mD_fD}b, the decay constant $ f_D a $ is plotted 
versus bare quark mass $ m_q a $. 
The data is well fitted by the straight line  
$ f_{D} a = 0.105(1) + 0.172(1)  \times  (m_q a) $. 
At $ m_q a = 0 $, it gives $ f_D = 235(8) $ MeV, which serves 
as a prediction of lattice QCD with exact chiral symmetry. 

The pseudoscalar meson of $ \bar s c $ or $ s \bar c $ 
corresponds to $ m_Q a = m_c a = 0.80 $ and $ m_q a = m_s a = 0.08 $. 
Its mass and decay constant are extracted directly from the 
time-correlation function, which are plotted as the eleventh 
data point (counting from the smallest one) in Fig. \ref{fig:mD_fD}.
The results are $ m_{D_s} a = 0.878(2) $ and $ f_{D_s} a = 0.119(2) $.  
The mass gives $ m_{D_s} = 1964(5) $ MeV, in good agreement with the 
mass of $ D_s(1968) $. The decay constant gives $ f_{D_s} = 266(10) $ MeV,
which agrees with the value $ f_{D_s^+} = 267 \pm 33 $ MeV  
complied by PDG \cite{Eidelman:2004wy}. 

\section{Ratios $ f_K/f_\pi $ and $ f_{D_s}/f_D $}

\begin{figure}[th]
\begin{center}
\begin{tabular}{@{}c@{}c@{ }c@{}c@{}}
\parbox[b][7.5cm][t]{6.5mm}{(a)} &
\parbox[b][7.5cm][t]{6.5cm}{
        \includegraphics*[height=7cm,width=6cm]{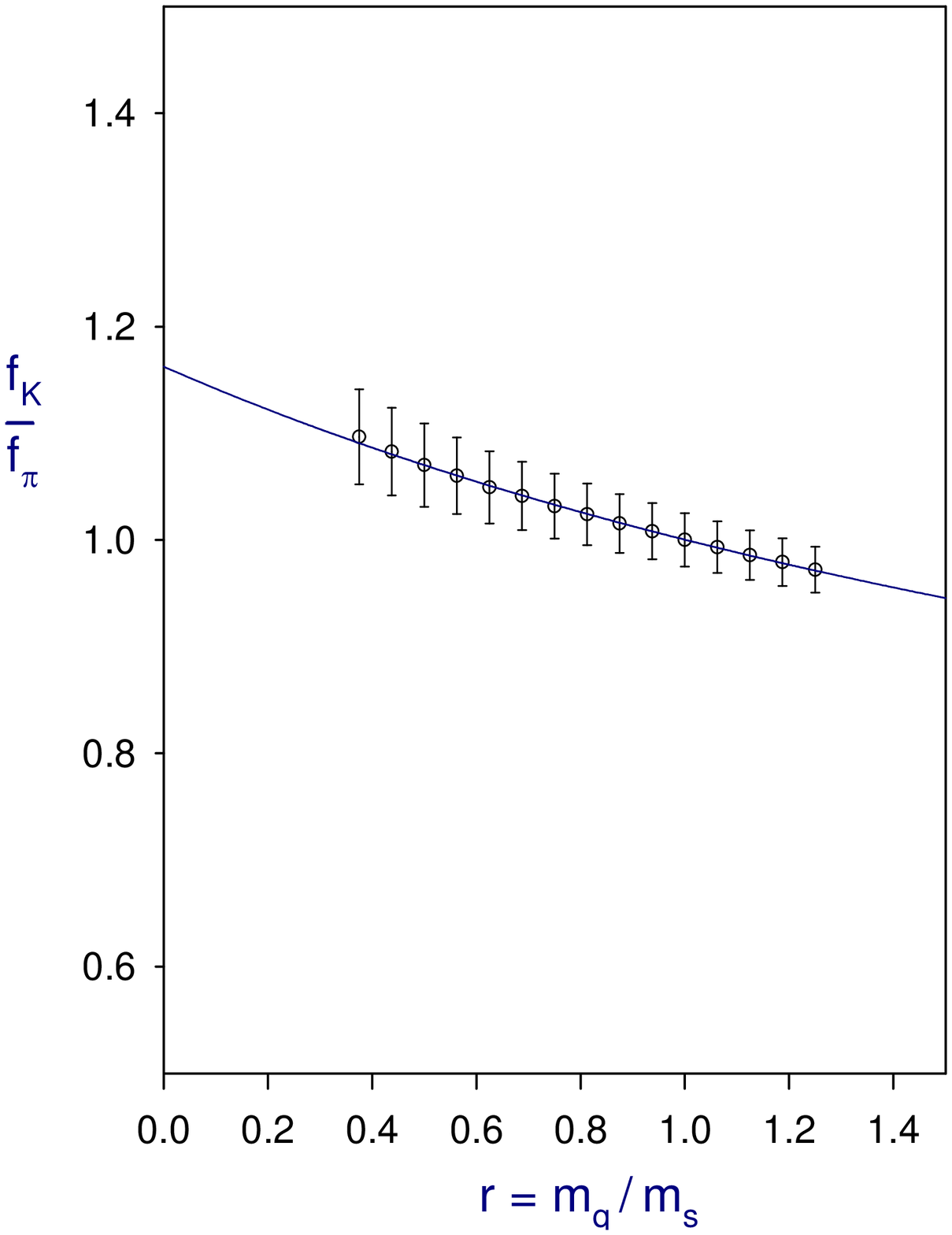}} &
\parbox[b][7.5cm][t]{6.5mm}{(b)} &
\parbox[b][7.5cm][t]{6.5cm}{
        \includegraphics*[height=7cm,width=6cm]{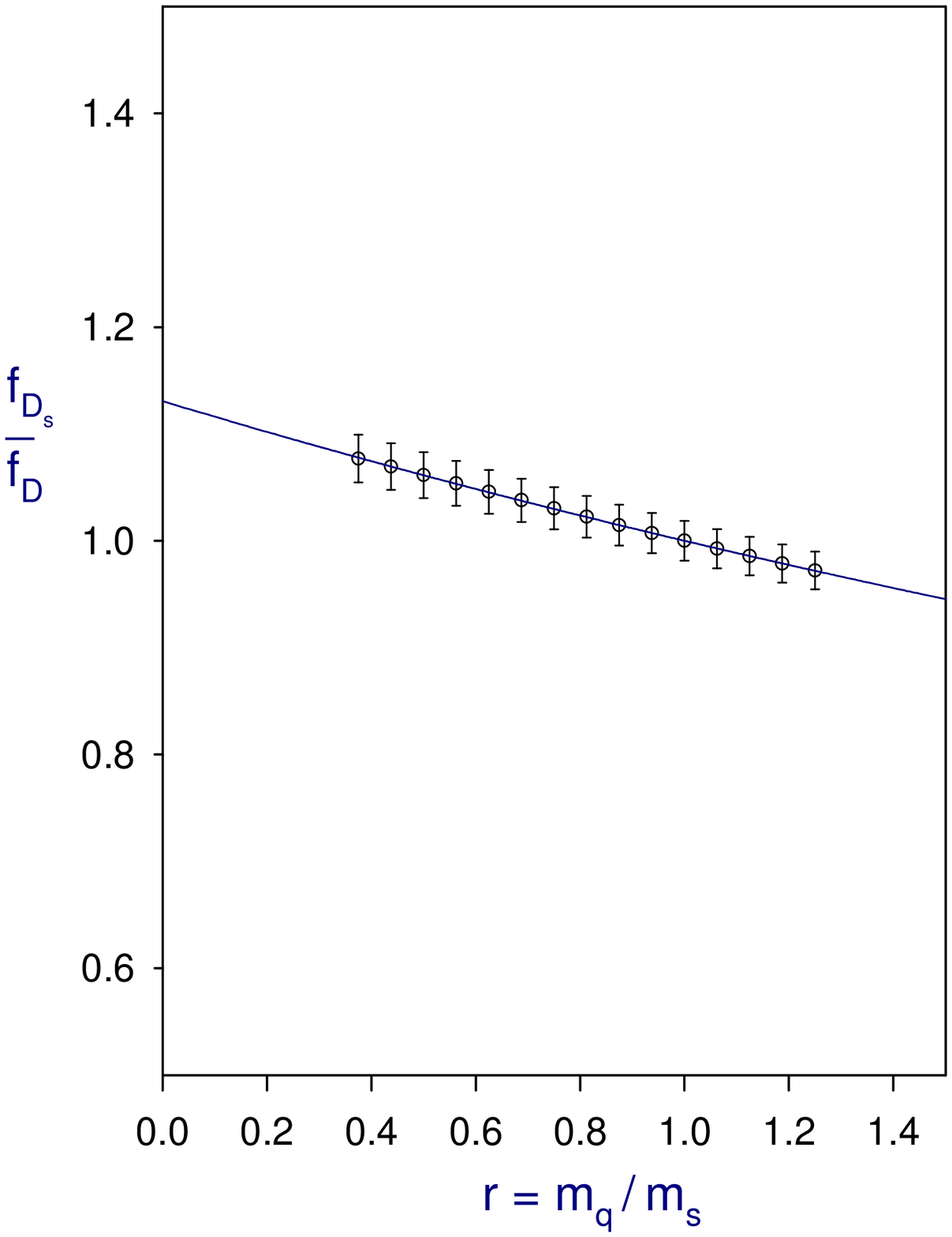}}
\end{tabular}
\vskip -0.8truecm
\caption{
(a) The ratio $ f_K/f_\pi$ versus $ r = m_u/m_s $
for 15 bare quark masses in the range $ 0.03 \le m_q a \le 0.10 $.
The solid line is the fit (4.1). 
(b) The ratio $ f_{D_s}/f_D $ versus $ m_u/m_s $. 
The solid line is the fit (4.2).}
\label{fig:rK_rD}
\end{center}
\end{figure}

At this point, it is instructive to obtain the ratios $ f_K/f_\pi $ 
and $ f_{D_s}/f_D $, and to see how well they agree the experimental values. 
In Fig. \ref{fig:rK_rD}a, the ratio $ f_{K^+}/f_\pi $ is plotted versus
$ r = m_u /m_s $ for 15 bare quark masses in the range 
$ 0.03 \le m_q a \le 0.10 $.
The data of $ f_{K^+}/f_\pi $ can be fitted by
\bea
\label{eq:rK_fit}
0.669(22) + \frac{0.931(116) }{1.813(163) + r}
\eea
At physical limit $ m_u/m_s = 1/26 $, 
it gives $ f_{K^+}/f_\pi = 1.17(6) $, in good agreement with 
the experiemntal value $ \sim 1.22 $ (PDG) \cite{Eidelman:2004wy}. 
With $ f_\pi = 131 $ MeV as the input, it gives $ f_{K^+} = 153(8) $ MeV, 
in good agreement with the experimental value   
$ f_{K^+} = 159.8 \pm 1.4 \pm 0.44 $ MeV 
complied by PDG \cite{Eidelman:2004wy}. 

Similarly, in Fig. \ref{fig:rK_rD}b,  
the data of $ f_{D_s^+}/f_{D^+} $ can be fitted by
\bea
\label{eq:rD_fit}
\frac{8.658(27) }{7.657(26) + r}
\eea
At physical limit $ r = 1/26 $, it gives $ f_{D_s^+}/f_{D^+} = 1.13(2) $,  
which serves as a theoretical prediction.

\section{Summary and Concluding Remarks}

In Ref. \cite{Chiu:2005ue}, 
we have determined the masses and decay constants 
of pseudoscalar mesons $ K $, $ D $ and $ D_s $, 
in quenched lattice QCD with exact chiral symmetry. 
Our results are:
\BAN 
& & m_K = 478 \pm 16 \pm 20 \mbox{ MeV},  \hspace{2mm}  
    m_D = 1842 \pm 15 \pm 21 \mbox{ MeV}, \hspace{2mm}  
    m_{D_s} = 1964 \pm 5 \pm 10 \mbox{ MeV},\\   
& & f_K = 152 \pm 6 \pm 10 \mbox{ MeV}, \hspace{2mm}
    f_D = 235 \pm 8 \pm 14 \mbox{ MeV}, \hspace{2mm} 
    f_{D_s} = 266 \pm 10 \pm 18 \mbox{ MeV},  
\EAN
where in each case, the first error is statistical, while 
the second is our crude estimate of combined systematic uncertainty.
Note that at the time when our results were posted at arXiv:hep-ph/0506266 
on June 26, the preliminary result of $ f_{D^+} $ from CLEO-c 
was $ f_{D^+} = 202(41)(17) $ MeV \cite{Bonvicini:2004gv}.
So our value of $ f_D $ was not in good agreement with the 
preliminary result of CLEO.  
However, with higher statistics at CLEO-c, the 
experimental value of $ f_{D^+} $ turns out to be 
$f_{D^+}=(222.6\pm 16.7^{+2.8}_{-3.4})~{\rm MeV} $ \cite{Artuso:2005ym},  
in good agreement with our prediction of $ f_D $.  
Now it remains to see whether the value of $ f_{D_s} $ 
coming from the high-statistics charm program of CLEO-c  
would agree with our value determined by lattice QCD with exact chiral 
symmetry. 

A salient feature of our calculation is that all quarks 
(no matter heavy or light) are treated on an equal footing,  
and they are fully relativistic with exact chiral symmetry on the lattice. 
Also, we have not used any heavy quark approximations for the charm quark. 
Even though our results are obtained in the quenched 
approximation, we suspect that for lattice QCD with exact chiral symmetry, 
the quenching error in $ f_D $, and $ f_{D_s} $ is 
less than $ 5\%$, in view of the good agreement between 
our result of $ f_K $ and the experimental value.  

In closing, we note that there are several quenched/unquenched
lattice QCD results for $ f_D $ and $ f_{D_s} $ in the literature
\cite{Becirevic:1998ua,Lellouch:2000tw,Aubin:2005ar}, 
as well as reported in this conference \cite{Simone:2005,Dong:2005}.
For the unquenched staggered quark with 
$ n_f = 2 + 1 $ \cite{Aubin:2005ar,Simone:2005}, 
its prediction of $f_{D^+}=(201 \pm 3 \pm 17)~{\rm MeV} $  
only agrees with the latest CLEO experimental result  
\cite{Artuso:2005ym} at $ 45\% $ confidence level, and is $ 10\%$ 
smaller than the experimental result.  
Note that this unquenched lattice QCD calculation relies on the  
so-called fourth-root trick to reduce the four tastes of each 
staggered quark to one taste, in computing the fermion determinant.  
Since this scheme for staggered quark has not achieved a high-precision 
prediction for $ f_{D^+} $, this may be an indication of the failure of the 
fourth-root trick in practice.


\begin{thebibliography}{15}

%\cite{Chiu:2005ue}
\bibitem{Chiu:2005ue}
T.~W.~Chiu, T.~H.~Hsieh, J.~Y.~Lee, P.~H.~Liu and H.~J.~Chang,
%``{\it Pseudoscalar decay constants $ f_D $ and $ f_{D_s} $ 
%    in lattice QCD with exact chiral symmetry}'', 
  Phys.\ Lett.\ B {\bf 624}, 31 (2005)
  [hep-ph/0506266].
  %%CITATION = HEP-PH 0506266;%%

%\cite{Artuso:2005LP}
\bibitem{Artuso:2005LP}
  M.~Artuso,
%``{\it Charm Decays Within the Standard Model and Beyond}'',
  plenary talk given at 
22nd International Symposium on Lepton-Photon Interactions at High Energy, 
Uppsala, Sweden, 30 Jun - 5 Jul 2005. 
%(http://lp2005.tsl.uu.se/~lp2005/LP2005/programme/presentationer/1\_artuso\_leppho05.pdf) 

%\cite{Artuso:2005ym}
\bibitem{Artuso:2005ym}
  M.~Artuso {\it et al.}  [CLEO Collaboration],
%  ``{\it Improved measurement of $ B(D^+ \to \mu^+ \nu) $ 
%    and the pseudoscalar decay constant $ f_{D^+} $}'',
  hep-ex/0508057.
  %%CITATION = HEP-EX 0508057;%%


%\cite{Chiu:2002ir}
\bibitem{Chiu:2002ir}
T.~W.~Chiu,
%``{\it Optimal lattice domain-wall fermions}'',
Phys.\ Rev.\ Lett.\  {\bf 90}, 071601 (2003);
%[hep-lat/0209153];
%%CITATION = HEP-LAT 0209153;%%
%
%\cite{Chiu:2002kj}
%\bibitem{Chiu:2002kj}
%T.~W.~Chiu,
%``Locality of optimal lattice domain-wall fermions,''
Phys.\ Lett.\ B {\bf 552}, 97 (2003); 
%[hep-lat/0211032];
%%CITATION = HEP-LAT 0211032;%%
%
%\cite{Chiu:2003ir}
%\bibitem{Chiu:2003ir}
%T.~W.~Chiu,
%``Aspects of domain-wall fermions on the lattice,''
hep-lat/0303008;
%%CITATION = HEP-LAT 0303008;%%
%\cite{Chiu:2003bv}
%\bibitem{Chiu:2003bv}
%T.~W.~Chiu,
%``{\it Recent developments of domain-wall/overlap fermions for lattice QCD}'', 
Nucl.\ Phys.\ Proc.\ Suppl.\  {\bf 129}, 135 (2004).  
%[hep-lat/0310043].
%%CITATION = HEP-LAT 0310043;%%

%\cite{Eidelman:2004wy}
\bibitem{Eidelman:2004wy}
S.~Eidelman {\it et al.}  [Particle Data Group Collaboration],
%``Review of particle physics,''
Phys.\ Lett.\ B {\bf 592}, 1 (2004).
%%CITATION = PHLTA,B592,1;%%

%\cite{Bonvicini:2004gv}
\bibitem{Bonvicini:2004gv}
  G.~Bonvicini {\it et al.}  [CLEO Collaboration],
  %``Measuring B(D+ $\to$ mu+ nu) and the pseudoscalar decay constant f(D+),''
  Phys.\ Rev.\ D {\bf 70}, 112004 (2004)
%  [hep-ex/0411050].
  %%CITATION = HEP-EX 0411050;%%

%\cite{Becirevic:1998ua}
\bibitem{Becirevic:1998ua}
  D.~Becirevic, P.~Boucaud, J.~P.~Leroy, V.~Lubicz, G.~Martinelli, F.~Mescia and F.~Rapuano,
  %``Non-perturbatively improved heavy-light mesons: Masses and decay
  %constants,''
  Phys.\ Rev.\ D {\bf 60}, 074501 (1999)
%  [hep-lat/9811003].
  %%CITATION = HEP-LAT 9811003;%%

%\cite{Lellouch:2000tw}
\bibitem{Lellouch:2000tw}
  L.~Lellouch and C.~J.~D.~Lin  [UKQCD Collaboration],
  %``Standard model matrix elements for neutral B meson mixing and  associated
  %decay constants,''
  Phys.\ Rev.\ D {\bf 64}, 094501 (2001)
%  [hep-ph/0011086].
  %%CITATION = HEP-PH 0011086;%%

%\cite{Aubin:2005ar}
\bibitem{Aubin:2005ar}
  C.~Aubin {\it et al.},
  ``{\em Charmed meson decay constants in three-flavor lattice QCD}'', 
  hep-lat/0506030.
  %%CITATION = HEP-LAT 0506030;%%

%\cite{Simone:2005}
\bibitem{Simone:2005}
  J.~Simone,
  ``{\em The determination of decay constants $ f_D $ and $ f_{D_s} $ in three
     flavor lattice QCD}'',poster presented at Lattice 2005. 

%\cite{Dong:2005}
\bibitem{Dong:2005}
  S.-J.~Dong,
  ``{\em $D_s$ meson decay constant $ f_{D_s} $ from quenched lattice QCD
     with overlap fermions}'', 
  talk presented at Lattice 2005. 

\end{thebibliography}
\end{document}